\documentstyle[12pt]{article}

\def\be{\begin{equation}}
\def\ee{\end{equation}}
\def\ba{\begin{array}{c}}
\def\ea{\end{array}}

\def\ben{$$}
\def\een{$$}
\begin{document}

\titlepage
%\vspace*{2cm}

 \begin{center}{\Large \bf
 Spiked and
${\cal PT}-$symmetrized
 decadic potentials
 supporting
   elementary $N-$plets of bound states
 }\end{center}

\vspace{5mm}

 \begin{center}
Miloslav Znojil$^\dagger$\footnote{e-mail: znojil@ujf.cas.cz }
\vspace{3mm}

$\dagger$
 \'{U}stav jadern\'e fyziky AV \v{C}R,
250 68 \v{R}e\v{z}, Czech Republic

\end{center}

\vspace{5mm}

\section*{Abstract}

We show that the potential wells $V^{}(x) = x^{10} + a\,x^8 +
b\,x^6+{c}\,x^4 +{d}\,x^2+{f}/x^2$ with a central spike possess
arbitrary finite multiplets of elementary exact bound states.
Their strong asymptotic growth implies an ambiguity in their
${\cal PT}$ symmetrically generalized quantization (via complex
boundary conditions) but the three eligible recipes coincide at
our exceptional solutions.

 \vspace{9mm}

\noindent
 PACS 03.65.Ge

%\vspace{9mm}

% \begin{center}
%{\small \today, dekadi.tex file}
%\end{center}

\newpage

\section{Introduction}

\noindent

Realistic calculations in quantum physics and field theory are
often guided by a parallel study of a simplified
quantum-mechanical model in one dimension. A typical example may
be found, e.g., in ref. \cite{Milton} where a self-interacting
scalar field theory is being modified in non-perturbative manner.
It is underlined there that one has to pay due attention not only
to the interaction potential $V(x)$ itself but also to the
boundary conditions imposed upon solutions in infinity.

In general the latter conditions are not unique. One of the most
transparent explicit examples of their ambiguity has been offered
by Bender and Turbiner \cite{BT}. In essence, they considered a
partially solvable sextic potential $V(x) = x^6 - 3\,x^2$ and its
zero-energy ground-state wave function $\psi(x) = \exp (-x^4/4)$.
In this model it is obvious that once you start working in the
whole complex plane, $x= \varrho\,e^\varphi \in l\!\!\!C$, the
asymptotic normalizability of the wave function keeps satisfied
not only on the real line but, in general, within the four
different asymptotic wedges defined by the elementary equation
${\rm Re}\,(-x^4) < 0$. This gives the admissible angles
$\varphi \in S_k$, $k = 1, 2, 3, 4$ lying within the four
separate intervals,
  \be
  \begin{array}{l}
 S_1=  (-\pi/8, \pi/8), \ \ \ \ \ \  S_2=
 (3\,\pi/8, 5\,\pi/8),\\ S_3=
 (7\,\pi/8, 9\,\pi/8),\ \ \ \ \  S_4=
  (11\,\pi/8,13\, \pi/8)\ .
  \ea
 \label{schema}
 \ee
Schematically, the situation is depicted in Figure~1.  One
imagines that the eligible {\em complex} physical
coordinates can be arbitrary curves with ends which do not enter
the ``forbidden" asymptotic domains.  Thus, the current real line
starts at $\varphi_L \in S_3$  in the left infinity and ends at
$\varphi_R \in S_1$. Its slightly bizarre ``Wick-rotated"
alternative $S_4 + S_2$ has also been discussed in the literature
as a weird example of a system whose real spectrum is not bounded
below \cite{Tater}. The elementary zero-energy bound state itself
is {\em shared} by both these non-equivalent spectra.

From  a retrograde point of view the choice of the sextic $V(x)$
proved unfortunate since its menu (\ref{schema}) is not yet rich
enough. Further progress has only been achieved five years later
when Bender and Boettcher \cite{BB} came to the conclusion that a
privileged role must be played by the pairs of sectors with a
mirror left-right symmetry. This symmetry keeps the trace of its
origin in field theory and is called ${\cal PT }$ symmetry. In the
quantum mechanical context the Hamiltonians $H$ have to commute
with the product of ${\cal P}$ (parity) and ${\cal T}$ (complex
conjugation or time reflection) \cite{Bender}. It is currently
believed that this guarantees the reality of the spectrum for many
non-Hermitian complex Hamiltonians \cite{proofs}.

In the very special class of these models
$V(x)=x^2\,(ix)^{2\,\delta} $ the complex plane is to be cut
upwards in order to keep the picture unique. Bender and Boettcher
\cite{BB} started from the smallest exponents $\delta$ and picked
up the lower sectors
 \ben
S_L=(-3\,\Delta-\pi/2,-\Delta-\pi/2),
 \ \ \ \ \ \
S_R=(\Delta-\pi/2,\,3\,\Delta-\pi/2)
 \een
with the half-width $ \Delta = \pi/(4+2\,\delta)$. Of course, for
any $\delta > 1/2$ there emerges an alternative pair of sectors
 \ben
S_{L2}=(-5\,\Delta-\pi/2,-3\,\Delta-\pi/2),
 \ \ \ \ \ \
S_{R2}=(3\,\Delta-\pi/2,\,5\,\Delta-\pi/2)\
 \een
which has been used by Buslaev and Grecchi in their study of
asymptotically quartic potentials \cite{BG}. In general the latter
possibility remains compatible with the current real coordinates
in the interval of $\delta \in (1, 3)$.

For all the asymptotically power-law models the ambiguity of
quantization is an interesting phenomenon.  Numerically, this has
been documented by several studies which made use of the limiting
transition $\delta \to \infty$ \cite{Rafa} or of the absence of
the cut at $\delta= 1, 2, \ldots$ \cite{bsqw}. An even simpler
form of the $\delta$ dependence takes place at the positive
integers $Z = 1+\delta/2$ in the potentials $V(x) = x^{4Z-2} +
{\cal O} (x^{4Z-3})$ since, asymptotically, their wave functions
$\psi(x) \approx \exp (-x^{2Z}/2Z)$ are symmetric on the real
line.

There appear the two new ${\cal PT}$ symmetric pairs of sectors
$(S_L, S_R)$ at each odd value of the integer $Z$. Thus, the
$\delta = 4$ and $Z=3$ decadic oscillator is the first model with
an ambiguity of this type. This is illustrated in Figure 2 where
the single left-right pair of the asymptotic boundary conditions
of Figure 1 is replaced by the triple choice.  All the three
angles $\varphi_L\in S_{Lj}$ and $\varphi_R \in S_{Rj}$ with $j =
1, 2$ or $3$ are equally compatible with the normalizability of
the exponential $\psi(x) \approx \exp (-x^{6}/6)$. With this
motivation we shall consider here all the decadic polynomial
potentials complemented by a central spike,
  \be V^{}(r) = r^{10} + a\,r^8 +
b\,r^6+{c}\,r^4 +{d}\,r^2 + {f}/r^2\ .
 \label{dekadicky}
 \ee
These forces contain as many as five independent coupling
constants and their real forms have been studied by several
authors in the literature \cite{decadici}.  Here, we shall
re-consider these ``spiked decadic" interactions within the
generalized quantum mechanics of Bender et al \cite{Bender}. It in
effect weakens the current Hermiticity of the Hamiltonian to its
mere ${\cal PT}$ symmetry. For this reason one can construct more
solutions in principle. We shall see below that such an
expectation is well founded, indeed.

\section{Decadic oscillators}

The results of study of Hermitian Hamiltonians with interactions
(\ref{dekadicky}) were summarized by Ushveridze in sec. 2.4 of his
monograph \cite{Ushveridze} on the partially (so called
quasi-exactly) solvable models. This summary implies that the
Hermitian decadic force lies somewhere in between the numerous
purely numerical models (for which ``it seems absolutely
unrealistic to find an exact solution" \cite{Ushveridze}) and
quasi-exactly solvable models in the narrower sense (where one
requires the algebraic solvability for a multiplet of K states at
any pre-determined integer K).

The latter category (say, type-I QES) lies already quite close to
the harmonic and other completely solvable models. Its properties
are best exemplified by the standard Hermitian sextic model (cf.
eq. (1.4.13) and related discussion in ref. \cite{Ushveridze}).
The former extreme without any solvability is usually illustrated
by the Hermitian quartic oscillator (cf. section 1.3 in ref.
\cite{Ushveridze}). In this comparison, the ``intermediate"
decadic models (e.g., eq. (2.4.5) in the Ushveridze's book) admit
merely a few ($K$) exact bound states with a strongly limited
multiplicity $K \leq 5$ \cite{decadici}. Such a property (let's
call it QES of type II) is entirely trivial at $K=1$ and still
quite easily achieved at the first few $K > 1$ \cite{Leach}.

\subsection{${\cal PT}$ symmetric regularization}

The best illustration of influence of the replacement of
Hermiticity by the mere ${\cal PT}$ symmetry of the Hamiltonian is
provided by the popular quartic oscillators. In ref. \cite{BBjpa}
these potentials were shown to belong to the type I of the QES
category. Obviously, ${\cal PT}$ symmetry plays a crucial role in
such an improvement of solvability.

The picture will be completed in what follows.  We are going to
demonstrate that a ${\cal PT}$ symmetrization of eq.
(\ref{dekadicky}) leads still to the maximal, type-I form of the
QES property.  We shall see that for the complexified decadic
oscillators the integer K may be chosen arbitrarily large. It is
worth noting that the same enhanced solvability admitting the
arbitrarily large multiplets remains freely applicable in any
spatial dimension~$D$.

In a preparatory step of the explicit constructions let us remind
the reader that in any ( = $\ell$-th) partial-wave projection the
$D-$dimensional differential Schr\"{o}dinger equation with a
virtually arbitrary {\em regular} central potential reads
  \be
\left [ -\frac{d^2}{dr^2} + \frac{L(L+1)}{r^2} +V^{}(r) \right ]
\,\psi(r) = E\,\psi(r), \ \ \ \ \ \ \ \ L = \ell+(D-3)/2.
\label{SE}
 \ee
In a more general perspective, we can add a centrifugal-like spike
to the regular force.  This is known to preserve the exact
solvability of the spiked harmonic oscillator \cite{PTHO}. In a
close parallel, we do not need to change the notation too much,
re-defining only the angular momenta $L=L({f})$ in such a way that
 \be
 L(L+1) = {f} +\left (
\ell+\frac{D-3}{2} \right ) \, \left ( \ell+\frac{D-1}{2} \right
). \label{relax}
 \ee
As a consequence, equation (\ref{SE}) may be assigned a pair of
independent solutions with the well known behaviour near the
central singularity $\sim 1/r^2$,
  \ben
\psi_1(r) \sim r^{-L(f)}, \ \ \ \ \ \psi_2(r) \sim r^{L(f)+1}.
 \een
In such a setting the ``forgotten possibility" of a suitable
ansatz lies in the {\em simultaneous} use of both these solutions
in the (complex) vicinity of the origin,
  \be
\psi(r) \sim {\cal C}_1 r^{-L}\,[1+{\cal O}(r^2)] +
{\cal C}_2
r^{L+1}\,[1+{\cal O}(r^2)]\ . \label{inspira}
 \ee
Similar idea is slightly counterintuitive but it has already been
used in several papers on the quartic oscillators with $\delta=1$.
In 1993, Buslaev and Grecchi \cite{BG} paved the way by the
mathematically rigorous example of introduction of the complex
coordinates for a problem of the present type. They achieved a
regularization of the centrifugal spike by a constant imaginary
shift of the real axis,
 \be
r = r(x) = x - i\,\varepsilon, \ \ \ \ \ \ \ \ x \in
(-\infty,\infty), \ \ \ \ \ \varepsilon > 0. \label{shift}
 \ee
This makes the centrifugal spikes smooth and fully regular at all
$x$,
 \ben
\frac{L(L+1)}{(x-i\,\varepsilon)^2}
=
\frac{L(L+1)(x+i\,\varepsilon)^2}{(x^2+\varepsilon^2)^2} =
-\frac{L(L+1)}{\varepsilon^2} + {\cal O}(x^2).
 \een
A longer discussion of some consequences of this type of the
${\cal PT}$ symmetric regularization has been provided by ref.
\cite{PTHO}. There, both the even and odd wave functions of the
exactly solvable harmonic oscillator and other models were
assigned their separate analytic continuations to the complex $x$.
In accord with expectations, the spectrum of the energies remained
real.

In the present paper we shall try to follow the same pattern and
imagine that the {\em two-term} ansatz (\ref{inspira}) may remain
compatible with the {\em single} Taylor-series expansion of
$\psi(r)$ (say, in the powers of $r^2$) {\em whenever} the ratio
$r^{L+1}/r^{-L}$ is itself equal to an integer power of $r^2$. In
the other words, under a suitable convention $ L+1>-l$, our key
assumption will read $L+1/2=M$ where $M$ can only be a positive
integer, $M = 1, 2, \ldots$.

We shall see below that this type of constraint will lead to
significant simplifications of the solutions as well as to their
easier interpretation. Indeed, the trivial choice of $f=0$ implies
that $M=\ell-1+D/2=1$. This mimics the four-dimensional $s-$wave
or two-dimensional $p-$wave situation since, in both cases, $L =
1/2$. Similarly we arrive at the alternative choice between the
six-dimensional $s-$wave, four-dimensional $p-$wave or
two-dimensional $d-$wave Schr\"{o}dinger equation at $M=2$, etc.

\subsection{Correct asymptotics and recurrences}

In 1998, Bender and Boettcher \cite{BBjpa} discovered the partial
solvability of the ${\cal PT}$ symmetrized quartic polynomial
oscillators. In the present language this means that they just
employed the ansatz of the type (\ref{inspira}) at the particular
$L = 0$. This has been accompanied by the asymptotically bent
choice of the complex integration path $r\to r(x) \in |\!\!\!C$.
In a way inspired by ref. \cite{BG} the latter construction has
been extended to all $L$ in our recent remark \cite{zobb}.

It is amusing to summarize that except the pioneering sextic study
\cite{BT} and its harmonic-oscillator simplification
\cite{BB,PTHO}, virtually all the available papers on the
(partially) solvable ${\cal PT}$ symmetric polynomial potentials
pay an exclusive attention to the asymmetric models $V(r)\neq
V(-r)$ of degree two \cite{is} and four \cite{BG,BBjpa,zobb}. In
this way all of them avoid the ambiguity problem but necessitate
the complex couplings and acquire the counter-intuitive property
${\rm Im}\ V(t)\neq 0$ even on the real axis of coordinates $t$.

The unpleasant asymmetries disappear within the decadic model
(\ref{dekadicky}) which is such that $V(r)= V(-r)$. Its asymptotic
growth is also steeper than in the current solvable models
\cite{JMP}. In what follows, we are going to show that this model
represents in fact the ``missing" last item in a list of all the
quasi-exactly solvable polynomial models. The first step towards
this not quite predictable result lies in the manifestly
normalizable ansatz
  \be
\psi(r) = \exp \left ( -\frac{r^6}{6} -\alpha\,\frac{r^4}{4} -
\beta\, \frac{r^2}{2} \right ) \,\sum_{n=0}^{N-1}\,h_n\,r^{2n-L}.
\label{ansatz}
 \ee
The insertion of this formula converts our differential
Schr\"{o}dinger equation (\ref{SE}) into the {\em finite} set of
recurrences
 \be
A_n\,h_{n+1} + B_n\,h_{n} + C_n\,h_{n-1} + D_n\,h_{n-2} =0, \ \ \
 \ \ \ \ \ n = 0, 1, \ldots, N\ . \label{recurrences}
  \ee
The use of the asymptotically optimal WKB-inspired parameters
 \ben
a = 2\alpha,
 \ \ \ \ \ \ \ \
b = \alpha^2+2 \beta, \ \ \ \ \ \ \ \ {c}={c}(N)=2\alpha
\beta+2M -4N-2
 \een
enables us to simplify the coefficients significantly,
 \be
\ba A_n=(2n+2)(2n+2-2M),\ \ \ \ \ \ B_n=E- \beta\,(4n+2-2M),\\
C_n=  \beta^2-{d} - \alpha\,(4n-2M) ,
 \ \ \ \ \ \ \
D_n=4\,(N+1-n). \ea \label{sear}
 \ee
This is the concise, linear algebraic formulation of our present
problem.

\section{Terminating solutions}

Centrifugal spike in eq. (\ref{SE}) binds the integers $M$ to its
strength ${f} \neq 0$ in a way prescribed by formula
(\ref{relax}). This means that certain non-vanishing spikes can
emerge as the simple functions of the dimension and angular
momentum,
 \ben {f}=M^2-(\ell-1+D/2)^2.
 \een
In particular, the most elementary choice of $D=M=1$ and $\ell=0$
implies that we just have to fix ${f} = 3/4$. This value is, by
the way, precisely equal to a boundary of a certain mathematical
regularity domain (cf. our recent remark \cite{PRA} for more
details).

\subsection{Sturmian multiplets of couplings at $M=1$}

We may notice that in accord with our definitions (\ref{sear}) two
coefficients vanish completely, $A_{-1}=0$ and $D_{N+1}=0$. This
is vital for the consistency of our ansatz (\ref{ansatz}). The
{third} consequence of our assumptions reads $A_{M-1}=0$ and has
no obvious interpretation. Seemingly, it is redundant.  Let us now
pay more attention to its crucial and beneficial role.

Starting from the first nontrivial choice of $M=1$ we get simply
$A_0=0$. In such a case we can fix $E = 0$ and discover that the
whole over-determined set of our recurrences (\ref{recurrences})
degenerates to the mere square-matrix recipe. As long as
$C_n=C_n({d})= \beta^2 - \alpha\,(4n-2M) - {d}$ we can write
  \be
 \left(
 \begin{array}{ccccc}
 C_1(0)&B_1&A_1&&\\
   D_2&C_2(0)&\ddots&\ddots&\\
    &\ddots&\ddots&B_{N-2}&A_{N-2}\\
 &&D_{N-1}&C_{N-1}(0)&B_{N-1}\\
 &
&&D_{N}&C_{N}(0)
 \ea \right )
\left ( \ba h_0\\ h_1\\ h_2\\ \vdots \\ h_{N-1} \ea \right ) =\
{d} \cdot \left ( \ba h_0\\ h_1\\ h_2\\ \vdots \\ h_{N-1} \ea
\right ).
  \label{matrix}
 \ee
We see that a more or less routine diagonalization of a
four-diagonal matrix determines in principle the $N$ different
eigen-couplings ${d} = {d}_k$ with $k = 1, 2, \ldots, N$.  These
values may be found numerically at an arbitrary $N$.  In the other
words, the whole problem becomes quasi-exactly solvable of type I.
This is our first important result.

It makes sense to introduce a shifted coupling $F={d}-
\beta^2+2N\,\alpha$. For the first few smallest dimensions $N \geq
0$ this simplifies the formulae and leads to the transparent
implicit definitions of the shifted couplings $F({d},N)$,
 \ben \ba
 F = 0, \ \ \ \ \ \ \ \ \ \ \ \ \ \ \ \ \ \ \ \ \ \ N=1,\\
{F}^{2}+16\,{ \beta}-4\,{{\alpha}}^{2}=0, \ \ \ \ \ \ \ \ \ \ \ \
N = 2,\\ -{F}^{3}+\left (16\,{{\alpha}}^{2}-64\,{ \beta}\right
)F+256=0 , \ \ \ \ \ \ \ \ N=3,\\ {F}^{4}+\left (160\,{
\beta}-40\,{{\alpha}}^{2}\right ) {F}^{2}-1536\,F
+144\,{{\alpha}}^{4}-1152\,{ \beta}\,{{\alpha}}^{2} +2304\,{{
\beta}}^{ 2}=0, \  \ \ \ N = 4,\\ -{F}^{5}+\left
(80\,{{\alpha}}^{2}-320\,{ \beta}\right ) {F}^{3}+5376\,
{F}^{2}+\left (8192\,{
\beta}\,{{\alpha}}^{2}-1024\,{{\alpha}}^{4}- 16384\,{{
\beta}}^{2}\right )F\\ +196608\,{ \beta}-49152\,{{\alpha}}^{2} =0,
\ \ \ \ \ N=5,\\ \ldots \ea
 \een We may summarize that
 at $M=1$,
our solutions remain purely non-numerical up to the degree $N =
4$.
  In a sufficiently broad
part of the $(\alpha, \beta)-$ plane (or, if you wish, of the
$(a,b)-$ plane of the octic and sextic coupling constants) the
multiplets of exact and elementary bound states exist and are
numbered by their (real) quadratic couplings $d_n$ at any value of
the multiplicity $N$.

\subsection{Multiplets of energies at $M=2$}

We have seen that our $M=1$ multiplets have been formed by
 the so called Sturmian solutions.
The role of their
energy was marginal and fixed to the single value
$E=0$.
Returning now back to our main story we
have to move to the
 next integer
$M=2$.
This
replaces the above-mentioned choice of
 $E=0$
(i.e., in effect, of $B_0=0$)
by the
virtually equally
efficient
simplification
  \ben \det \left(
 \begin{array}{cc}
B_0&A_0\\
 C_1&B_1
 \ea \right )
=0
  \label{4.8}
 \een
which leads immediately to the compact formula
 \ben
{d}={d}(E)=\frac{E^2}{4}, \ \ \ \ \ M=2.
\een
The insertion of this
energy-dependent
harmonic strength
$d$ returns us back to the
square-matrix
secular equation
(\ref{matrix}). With
its three innovated diagonals
 \ben
A_n=(2n+2)(2n-2),\ \ \ \ B_n=E- \beta\,(4n-2),
 \ \ \ \  \
C_n=  \beta^2-E^2/4 - \alpha\,(4n-4) \een it defines the spectrum
$E_n$ in the very similar manner as above, i.e., as roots of a
certain polynomial. We can display its first nontrivial
sample, $$ \ba -{{ E}}^{5}-2\,{ \beta}\,{{ E}}^{4}
 +\left (-48\,{\alpha}+8\,{{
 \beta}}^{2}\right ){{ E}}^{3}
+\left (-96\,{ \beta}\,{\alpha}+192+ 16\,{{ \beta}}^{3}\right ){{
E}}^{2}\\ +\left (-256\,{ \beta}-512\,{{ \alpha}}^{2}+192\,{{
\beta}}^{2}{\alpha}-16\,{{ \beta}}^{4}\right ){
 E}\\
-1024\,{ \beta}\,{{\alpha}}^{2}-1280\,{{ \beta}}^{2}+4096\,{
\alpha}-32\,{{ \beta}}^{5}+384\,{{ \beta}}^{3}{\alpha} = 0,\\ M=2,
\ N = 3. \ea $$ Its roots are not non-numerical anymore but three
of them remain manifestly real at $\alpha= \beta=0$ where their
form remains closed, $(E_1, E_2, E_3) = (0, 0, \sqrt[3]{192})$.
The consequent and precise specification of the whole domain of
the reality of these roots remains as numerical a task as, say,
its parallel studied in ref. \cite{BBjpa}.

\subsection{More complicated multiplets at $M = 3$ etc.}

Explicit formulae with the integers $M \geq 3$ become appreciably
more complicated. For all the really large truncations  $N \gg M$
they may remain useful. Their derivation would proceed along the
same lines as before. One starts from the general pre-conditioning
requirement
  \be \det \left(
 \begin{array}{ccccc}
B_0&A_0&&&\\
 C_1&B_1&A_1&&\\
   D_2&C_2&\ddots&\ddots&\\
    &\ddots&\ddots&B_{M-2}&A_{M-2}\\
 &&D_{M-1}&C_{M-1}&B_{M-1}
 \ea \right )
=0
  \label{eq4.8}
 \ee
and its solutions have to be inserted in the, presumably, much
larger main secular determinant
  \be
\det
 \left(
 \begin{array}{ccccc}
 C_1&B_1&A_1&&\\
   D_2&C_2&\ddots&\ddots&\\
    &\ddots&\ddots&B_{N-2}&A_{N-2}\\
 &&D_{N-1}&C_{N-1}&B_{N-1}\\
 &
&&D_{N}&C_{N}
 \ea \right )
=0\ .
 \label{detik}
 \ee
The procedure only becomes inefficient at the larger integers $M$.
In such a setting it would be more appropriate to treat both the
``small" and ``large" secular equations (\ref{eq4.8}) and
(\ref{detik}) on an equal footing, as a mutually coupled algebraic
system.

In the $M \approx N$ setting one suddenly loses the main advantage
of our present construction, viz., its reducibility to the single
secular equation. In practice, one also has to replace the
insertions of $E$ or $d$ by the use of the so called Gr\"{o}bner
bases. The algorithm  determines both $d$ and $E$ at once and is
routinely provided by the languages like MAPLE \cite{Maple}.

Although the generalized $M > 3$ formulae and calculations need
not necessarily become hopelessly complicated (and, in fact, give
the nice results, e.g., in the $M \to \infty$ limit \cite{pert})
their use would definitely require another motivation.  Once we
get that far, we would have no reason for maintaining our main
assumption (\ref{relax}).  At any real quantity $M=2L+1$
(reflecting a free variability of the coupling ${f}$) we would
only have to replace the ``small" dimension $M$ in eq.
(\ref{eq4.8}) by the overall truncation $N$ itself.

\section{Concluding remarks}

Results of our explicit constructions share their transparent
algebraic character: Mathematically, one eliminates the auxiliary
or ``redundant" root of ``the simple" condition (\ref{eq4.8}) and
ends up with the {\em single} ``effective-Hamiltonian-like"
square-matrix eigenvalue problem (\ref{detik}). This is the main
merit (and general feature) of the QES systems of type I. Indeed,
once we return to the explicit Hermitian decadic $N=2$
construction of sec. 2.4 in ref. \cite{Ushveridze}, we discover
that, in the same language, the main shortcoming of the type-II
QES constructions lies precisely in the necessity of solving {\em
several coupled} eigenvalue problems at once.

Our present sample non-numerical constructions available at the
first few smallest integers $M$ parallel in fact  many other
quasi-exactly solvable models. In particular, the Sturmian
constant-energy form of the elementary multiplets characterized
already the very first quasi-exactly solvable (viz., Coulomb plus
harmonic) model as discovered by A. Hautot in 1972 \cite{Hautot}.
More complicated relation between the couplings and energies
characterizes, e.g., the partially solvable anharmonicities $\sim
(1+g\,r^2)^{-1}$ revealed by G. Flessas in 1981 \cite{Flessas} and
explained algebraically in 1982 \cite{Whitehead}.

In our present paper the most elementary multiplets using
auxiliary integer $M=1$ proved purely Sturmianic. Their
spike-shaped short-range part of the interaction $\sim f$ is not
arbitrary and can only vanish in even dimensions. Vice versa, the
emergence of a spike for our $s-$wave multiplets in three
dimensions is reminiscent of the so called conditionally solvable
models where the choice of ${f} = 3/4$ is obligatory \cite{PRA}.

At any auxiliary $M$, in comparison with all the Hermitian QES-I
systems related to certain tridiagonal matrix representations of
Lie algebras \cite{Turbiner}, all the decadic examples are
distinguished by the four-diagonal and $N-$dimensional secular
determinants (\ref{detik}). In the Hermitian setting this
``four-diagonality" was in fact the main reason of the restricted
type-II solvability. We have shown that only the appropriate
complexification can move the decadic systems to a higher, type-I
QES group. At present, this group which exhibits the appealing
${\cal PT}$ symmetry already encompasses the asymptotically
decreasing quartic forces of Bender et al \cite{BBjpa,zobb} and
the Coulomb + harmonic model \cite{is}.

\section*{Acknowledgement}

Partially supported by the GA AS CR grant Nr. A 104 8004.

\section*{Figure captions}

\subsection*{
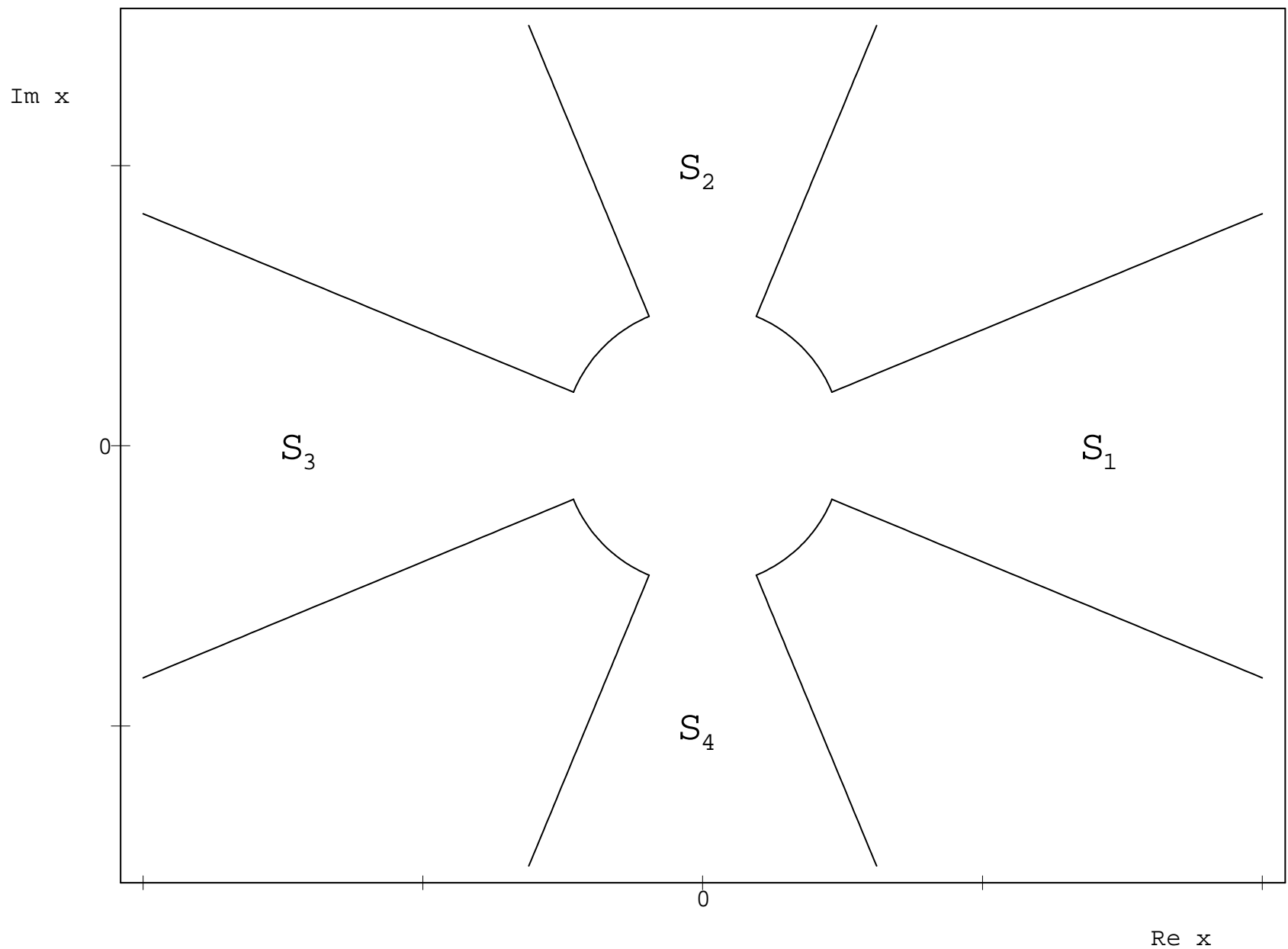
 Figure 1. Permitted domain for sextic oscillators}

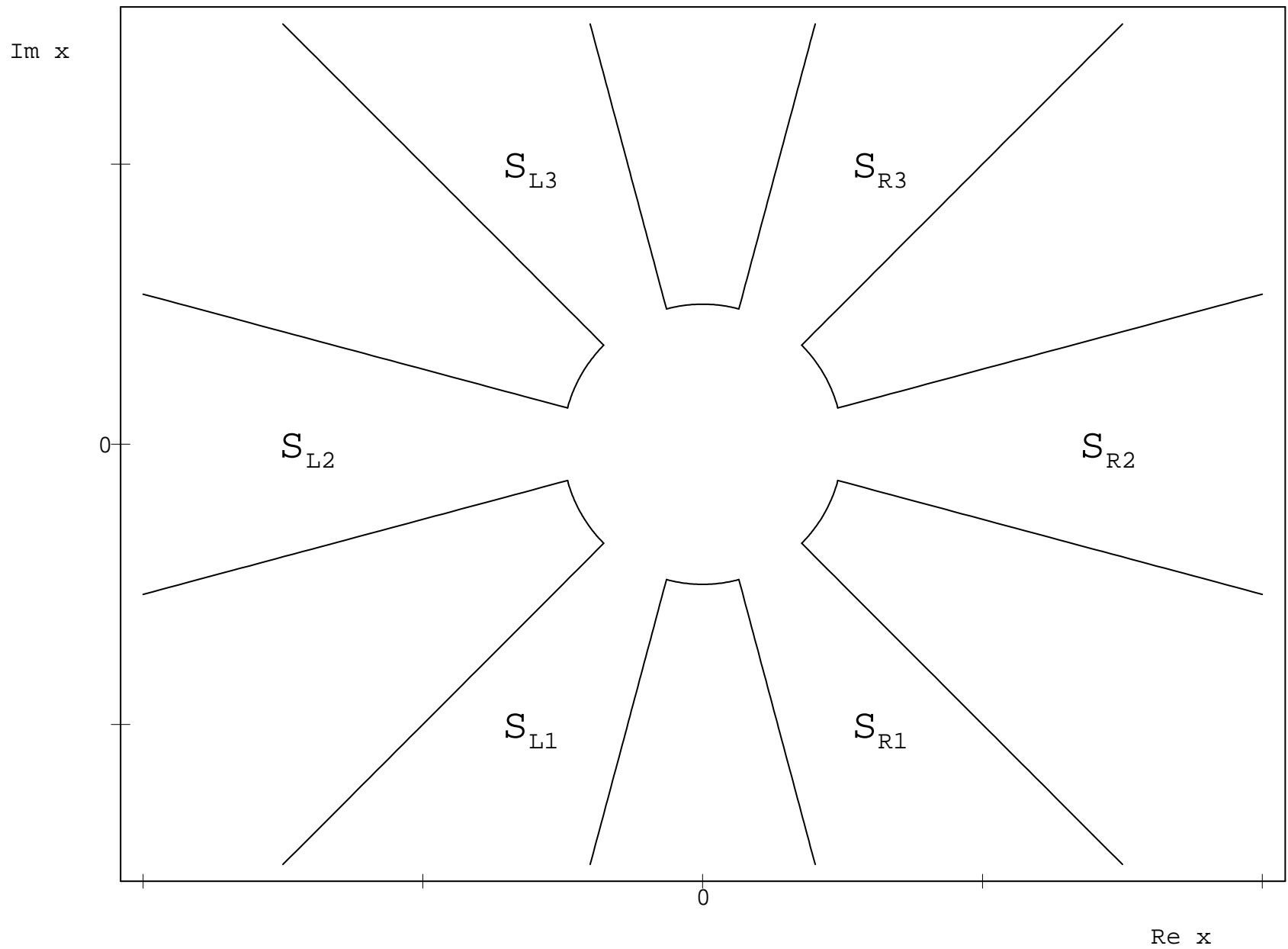
\subsection*{Figure 2. Permitted domain for decadic oscillators}

\end{document}